# Patterning of 2D second harmonic generation active arrays in ferroelectric nematic fluids.

M. Lovšin[1,2], A. Petelin[2], B. Berteloot[3], N. Osterman[1,2], S. Aya[4,5], M. Huang[4,5], I. Drevenšek-Olenik[1,2], R. J. Mandle[6,7], K. Neyts[3,8], A. Mertelj[1], N. Sebastian[1]

1 Jožef Stefan Institute, Ljubljana, Slovenia
2 University of Ljubljana, Faculty of Mathematics and Physics, Ljubljana, Slovenia
3 Liquid Crystals and Photonics Group, ELIS Department, Ghent University, Ghent, Belgium
4 South China Advanced Institute for Soft Matter Science and Technology (AISMST), School of Emergent Soft Matter, South China University of Technology, Guangzhou, China
5 Guangdong Provincial Key Laboratory of Functional and Intelligent Hybrid Materials and Devices, South China University of Technology, Guangzhou, China
6 School of Physics and Astronomy, University of Leeds, Leeds, UK
7 School of Chemistry, University of Leeds, Leeds, UK
8 State Key Lab of Advanced Displays and Optoelectronics Technologies, Hong Kong University of Science and Technology, Hong Kong

Corresponding author: nerea.sebastian@ijs.si

Ferroelectric nematic liquid crystals exhibit unique non-linear optical properties, with the potential to become transformative materials for photonic applications. A promising direction relies on the fabrication of tailored polar orientational patterns via photoalignment, thus shaping the non-linear optical susceptibility through thin slabs of the ferroelectric fluid. Here, we explore the fabrication of 2D periodic SHG active arrays in ferroelectric nematic fluids, for different materials, cell thicknesses and motifs. Based on polarizing optical microscopy observations in combination with optical simulations, second harmonic generation microscopy and interferometry, the 3D structure of the motifs are revealed. Two different 2D periodic patterns are explored, showing that the balance between flexoelectric and electrostatic energy can lead to different domain structures, an effect which is rooted in the difference between the flexoelectric properties of the materials. It is shown that by combining the surface-inscribed alignment with different spontaneous degrees of twist, 2D SHG active arrays can be obtained in the micrometre scale, in which adjacent areas exhibit maximum SHG signals at opposite angles.

## Introduction

The widespread implementation of nematic liquid crystals in optical devices results from a combination of their large optical anisotropy, fast reorientation under the application of small electric fields and the possibility of orientational control via confinement boundary conditions. The standard nematic phase is uniaxial and non-polar, i.e. centrosymmetric, and thus non-linear optical properties such as second harmonic generation (SHG) are only observed by breaking the inversion symmetry, for example by electric poling (electric field-induced SHG, i.e EFISHG) [1–3] or by the flexoelectric effect through spatial deformations [4]. It should be highlighted here, that in either case, SHG signals are very weak.

The discovery of a polar version of the nematic phase, in which molecules exhibit ferroelectric ordering [5–9], has been accompanied by a range of studies showing their potential for non-linear optical (NLO) devices. Large NLO coefficients have been reported [10,11] with the same order of magnitude as classical solid materials such as β-Barium borate (BBO). This enables a wide range of Second Harmonic Generation (SHG) microscopy studies, including the polar ordering and topology during nucleation of $N_F$ droplets at the direct isotropic to ferroelectric phase transition [12,13].

The localized tunability of the SHG signal by application of small fields was demonstrated in in-plane switching liquid crystal cells [14]. More recently, FNLCs have been demonstrated to be an efficient electric-field tunable broadband source of entangled photons based on spontaneous parametric down-conversion [15]. In the case of the cholesteric $N_F$ counterpart [16–19] ($N^*_F$), it has been shown that phase-matched SHG can be generated in the photonic bandgap [20,21].

The spontaneous polarization of FNLCs highly impacts their behaviour under confinement conditions, as in addition to the orientational coupling at the surfaces, polar effects are also present. In standard polyimide rubbed surfaces, it has been shown that, for materials such as RM734 and FNLC-919, polarization orients opposite to the rubbing direction [22,23]. Thus, antiparallel rubbed cells commonly used for homogeneous alignment in non-polar nematics, result in $\pi$-twisted structures across the cell thickness [24]. Interestingly, escape into twisted structures of both handedness has been reported also for unconstrained surfaces, suggesting that, to minimize electrostatic energy, the equilibrium state in thin layers is twisted [25].

We recently explored the alignment of FNLCs in patterned photoaligned cells, showing that flexoelectric coupling between splay deformation and polarization can be used to control the polarization direction [26]. From the designed periodic splay patterns, regions of alternating polarization direction were created in regions of opposite splay. Interestingly, at the edges of the patterned splay structures, where splay changed sign, the appearance of SHG inactive disclination lines separating the opposite polarization domains was described. In contrast to the non-polar high-temperature nematic phase in which the surface inscribed splay pattern was maintained across the cell thickness, it was shown that in the polar phase, the depolarization field created by bound charges $-\nabla \cdot \mathbf{P}$ caused the escape towards a uniform orientation in the middle of the cell, i.e. the "unsplay" of the structure [26].

The combination of NLO properties and the enabled possibility of designing custom polarization structures via patterned photoalignment is highly interesting for creating NLO active arrays, spatially modulating the NLO properties. In this contribution, we explore the fabrication of 2D SHG active arrays by photopatterning of pure 2D splay modulated structures and 2D splay-bend structures. For that purpose, we study such structures by combining polarizing optical microscopy, SHG microscopy and SHG interferometry experimental investigations with simulations of spectral transmission and second-harmonic generation (SHG) for thin birefringent media. Three different materials are employed: the archetypical DIO and RM734, in comparison with the 1D splay structures already reported for them [26] and FNLC-1571, a room-temperature ferroelectric nematic material.

## 2 Material and methods

### 2.1 Materials

Three different liquid crystalline materials have been employed: DIO [5], RM734 [7,27] and FNLC-1571. The latter was provided by Merck Electronics KGaA and exhibits three distinct thermotropic LC phases, with the ferroelectric nematic phase stable at room temperature: Isotropic- 88 °C – Nematic (N) - 62 °C – modulated antiferroelectric nematic - 48 °C – ferroelectric nematic phase ($N_F$) - 8 °C – crystal.

DIO material trans(2,3',4',5'-tetrafluoro-[1,1'-biphenyl]–4-yl 2,6-difluoro-4-(5-propyl-1,3-dioxan-2-yl)benzoate) has been synthesized according to the description given in reference [28]. On cooling, DIO exhibits the $N_F$ phase in the temperature interval 68.9 °C -60 °C on which we focused our investigations. Finally, RM734 (4-((4-nitrophenoxy) carbonyl)phenyl-2,4-dimethoxybenzoate) was synthesized according to reference [27]. Investigations of RM734 were performed in the $N_F$ phase accessible on cooling in the temperature range 132.7 °C -90 °C. Both materials have been extensively investigated [8,9,24,29,30].

### 2.2 Photopatterned cell preparation

Liquid crystal cells were built by assembling two glass substrates coated with indium tin oxide (ITO) and a layer of the photo-alignment material Brilliant Yellow separated by glass spacers of different $\mu m$-size diameters. The assembled thickness of the cells employed in this study was measured to be in the interval 2.8-10.1 $\mu m$. The liquid crystal orientation at the substrate surface was defined through patterned photoalignment, which provided anisotropic nonpolar in-plane alignment without pretilt. A detailed description of the photopatterning process can be found in references [26,31]. Each cell (2 x 2 cm$^2$) was illuminated in multiple areas producing an array of different photopatterned designs with dimension 1.3 x 0.73 mm$^2$. In between the patterns, the photo-alignment material was not illuminated, and thus the studied patterns are separated by regions with random planar alignment.

### 2.3. Polarizing optical microscopy and Berreman calculus.

POM experiments were performed in either a Nikon Eclipse or an Optiphot-2 POL Nikon microscope. Images and videos were recorded with a Canon EOS M200 camera. The sample was held in a heating stage (Instec HCS412W) together with a temperature controller (mK2000, Instec).

To calculate transmission spectra and color rendering images we used the diffractive transfer matrix method "dtmm" open software package [32], which uses the Berreman 4 x 4 matrix method. A detailed description of its implementation for the calculation of transmission spectra is given in the references [14,26] and in the online manual available at GitHub [32].

For SHG computation, we developed an iterative algorithm based on the technique presented in [33], where the authors use a modified Berreman calculus to compute the fundamental and the SHG waves assuming weak depletion of the fundamental beam. In short, we use the Berreman calculus to compute and obtain the electromagnetic field of the fundamental wave as it propagates through the sample. We calculate the gain coefficients for the SHG wave from the obtained fundamental beam at each layer of the optical stack and propagate the generated SHG wave through the stack. Compared to the technique in [33], our method works with weak and large depletion and has a more straightforward implementation. Details of the technique will be presented elsewhere. SHG simulations were performed for RM734 patterns taking into account the dispersion of the indexes of refraction $n_{2\omega} - n_{\omega}$ depicted in Fig.SI.1, determined as described in Supplementary Note I via the wedge-cell method. Nonlinear susceptibility coefficients $d_{33}$=5.6 pmV$^{-1}$ and $d_{31}$=0.05$d_{33}$ are considered [10].

## 2.4 Second Harmonic Generation Microscopy

Second Harmonic Generation microscopy (SHG-M) and interferometry (SHG-I) were performed using a custom-built sample-scanning microscope with an Erbium-doped fibre laser (C-Fibre A 780, MenloSystems, 785 nm, 95 fs pulses at a 100 MHz repetition rate) as the laser source. In order to avoid sample degradation, the average power was adjusted to 30 mW using an ND filter. A detailed description of the setup is available in reference [26]. The setup allows for the insertion of a BBO reference crystal before the sample followed by a Michelson interferometer for time compensation between the reference and the fundamental pulse. Additionally, the phase of the reference can be adjusted by fine rotation of a glass plate mounted on a motorized rotator to perform interferometric SHG imaging (SHG-I). The final SHG-M and SHG-I images are acquired with a high-performance CMOS camera (Grasshopper 3, Teledyne Flir) with a typical integration time of 250 ms, varying dimensions in pixels with a resolution of 0.285 µm/pixel. SHG-I interferograms of the areas of interest are then calculated by computing the mean intensity and the standard error of the mean intensity.

## 3. Results and Discussion

### 3.1 Pattern A: Splay 2D pattern in thin cells

We first explore 2D patterns as those depicted in Fig.1, in which in x-direction anchoring in both cell surfaces is shaped as $\mathbf{n}_s = (n_x, n_y) = (\sin(\vartheta_{surf}), \cos(\vartheta_{surf}))$, $\vartheta_{surf} = \vartheta_0 \sin(2\pi x/P)$, with $\vartheta_0$ = 40 degrees and the period P=$2\pi/k$= 40 µm. In the y-direction the modulation phase is shifted by $\pi$ every 40 µm. Such pattern implies a maximum splay curvature $k\vartheta_0$ of 0.1 µm$^{-1}$ equivalent to that studied previously for 1D patterns [26]. Cells of different thicknesses (targeted d=3, 5 and 8 µm) were prepared. At this stage, we explored the behaviour of DIO and RM734 in 3 µm cells. The quality of the alignment was first checked in the high temperature non-polar nematic (N) phase, showing good alignment across the whole area of the design. The modulation shift across the y-direction, results in the tessellation of surface line defects at discrete y-positions as shown in Fig.SI.2.

On cooling towards the N$_F$ phase, both materials show very different behaviour. Similarly, as reported for the 1D patterns, upon transition from the modulated antiferroelectric phase into the N$_F$ phase, DIO initially shows the homogenization of the observed stripped texture in the intermediate antiferroelectric phase together with the appearance of defect lines parallel to the positions at which splay changes sign (lines running along y-direction) (Fig.SI.3). Noteworthy, such defect lines also appear at an angle in many regions of the pattern, following the diagonal of the 2D design and heavily influenced by the surrounding frame's uniform direction. Interestingly, no new defect lines are observed running along x-direction in the positions at which the phase shift occurs. Such texture remains stable for around 4 degrees below the transition, before an additional macrostructural relaxation takes place propagating along the defect lines. On the other hand, the texture in the direct N-N$_F$ transition of RM734 rapidly homogenizes and although initially some faint defect lines can be distinguished (Fig.SI.4), they rapidly shrink and move out of the pattern, only remaining at a few locations near the edges of the area (Fig.S.4).

Fig. 1.a-b shows the full patterned area for DIO (before structural relaxation) and RM734 in the N$_F$ phase. In both cases two differentiated behaviours can be observed, showing symmetric or asymmetric optical behaviour under crossed polarizers. Qualitatively the director structure throughout the pattern and the cell thickness can be established by comparing POM observations and dtmm simulations. Let us first focus on the areas with symmetric behaviour (Fig.1.d&f), the fact that upon sample rotation, not only the intensity but also the spectral characteristics change across the pattern indicates that the prescribed alignment is not maintained in the z-direction (along the cell thickness). That is, equivalently to the simpler 1D patterns, while at the surfaces the director follows the photopatterned anchoring direction, and twists towards the cell centre in order to reduce the splay as $\vartheta(z) = \vartheta_{surf} e^{(2z^2/d^2 - 1/2)/\xi}$ with $\xi = 0.2$. This "unsplay" minimizes

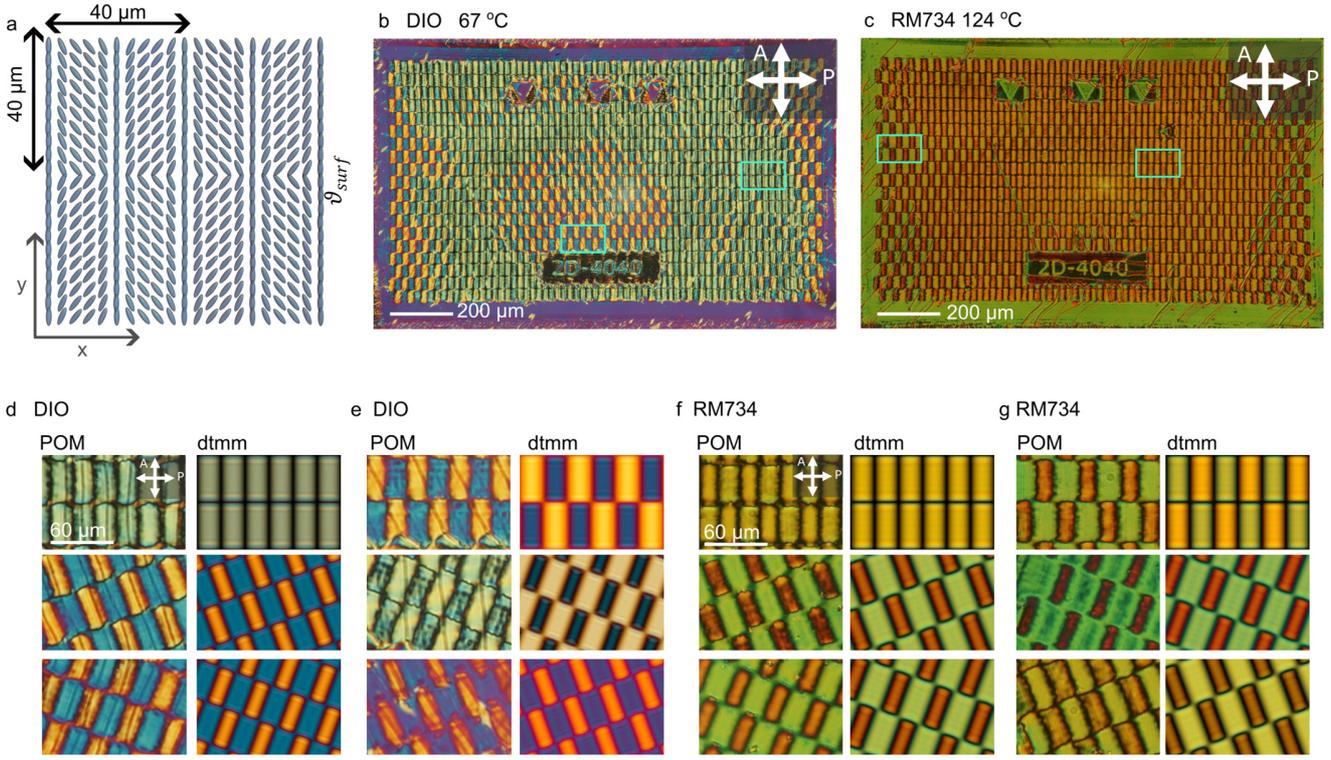

Fig.1. 2D pure splay patterns in DIO and RM734 in 3 μm cells. a) Photopatterned design inscribed in both cells' surfaces. b) POM image of the full patterned area for DIO at 67 ℃ under crossed polarizers. c) POM image of the full patterned area for RM734 at 124 ℃ under crossed polarizers. Patterns are surrounded by a frame with uniform alignment at 45 degrees from the vertical direction. Top arrows in the pattern indicate the x-positions at which alignment corresponds with vertical direction. d-g) Zoom-in images of the blue-framed regions in b and c. POM images are compared with dtmm transmission spectra simulations in regions with symmetric and asymmetric behaviour at extinction position and upon rotation. Double headed arrows indicate the direction of crossed polarizers. Dtmm simulations were performed considering Δn=0.19 for DIO and Δn=0.25 for RM734, where $n_o$ is taken to be 1.52.

bulk bound charges due to $-\nabla \cdot \mathbf{P}$ and as inferred from POM observations, it is common for both materials. In those regions with checkerboard colour pattern at the extinction position, the transmitted spectra are well reproduced, considering that in addition to the described "unsplay" the director twists so as in the middle of the cell the director is at an angle $\vartheta_{center}$ of around 30 degrees (corresponding to the diagonal direction for the half period in the x-direction and full period in the y-direction) according to:

$$\vartheta(z) = \vartheta_{surf} e^{\left(\frac{2z^2}{d^2}-\frac{1}{2}\right)/\xi} + \vartheta_{center}\left(1 - e^{\left(\frac{2z^2}{d^2}-\frac{1}{2}\right)/\xi}\right). \quad 1$$

The final equilibrium structure is the result of a fine interplay between elastic, flexoelectric and electrostatic energies. If we consider the structure seen for DIO, where periodic defect lines appear, the "unsplay" across the cell thickness, not only reduces bulk bound charges but also reduces the surface charge associated with the polarization change at the walls with surface charge density $\sigma_P = (\mathbf{P}_2 - \mathbf{P}_1) \cdot \mathbf{m}$, where $\mathbf{P}_{1,2}$ are the polarizations of the adjacent domains and $\mathbf{m}$ is the normal to the wall pointing into the domain with $\mathbf{P}_1$ (Fig.2.a). However, by the same argument, and provided that flexoelectric coupling would be enough in this motif to promote a 2D patterning of $\mathbf{P}$, the structure would need to be able to equilibrate a large number of 180° strongly charged domain walls periodically in y-direction (Fig.2.a). Such walls are not observed, indicating that the energy cost of adapting to unfavourable splay is smaller than the energy cost of the strongly charged walls. This is subsequently demonstrated by SHG interferometry measurements shown in Fig.2.b. The interferogram for DIO shows that adjacent opposite splay signs in the x-direction have opposite directions of $\mathbf{P}$, while the $\mathbf{P}$ direction is maintained in the y-direction.

Interestingly, the fast migration outside the pattern of defect lines at the transition in the case of RM734, as opposed to the behaviour in 1D splay design (Fig.SI.5), indicates that for this material a large monodomain is obtained, where the polarization adapts to the unfavourable splay in order to prevent the larger energy cost of forming domain walls. The flexoelectric term can be included in the free energy as $\frac{1}{2}K_1|\mathbf{S} - \mathbf{S_0}|^2$ where $\mathbf{S_0} = \gamma \mathbf{P} \cdot /K_1$, $\gamma$ is the bare splay flexoelectric coefficient, $K_1$ is the splay elastic constant and $\mathbf{n} \cdot \mathbf{S_0}$ is the ideal splay curvature, which would minimize the splay elastic energy. It was shown that patterning through flexoelectric coupling was possible for lower splay curvatures in the case of RM734 as compared with DIO [26], indicating a smaller ideal splay curvature in the case of the former. Thus, for comparable $K_1$ values, the energy cost of adapting to unfavourable splay should be lower in the case of RM734 than in the case of DIO, in agreement with the observations gathered in Fig.2. The structural difference between the areas with symmetric or asymmetric appearance can be attributed to the combined effect of localized asymmetries in the inscribed pattern and the minimization of electrostatic energy. Unsplaying towards an in-plane angle along the pattern diagonal causes areas of the same splay sign to connect, allowing to escape the need to adapt to unfavourable splay.

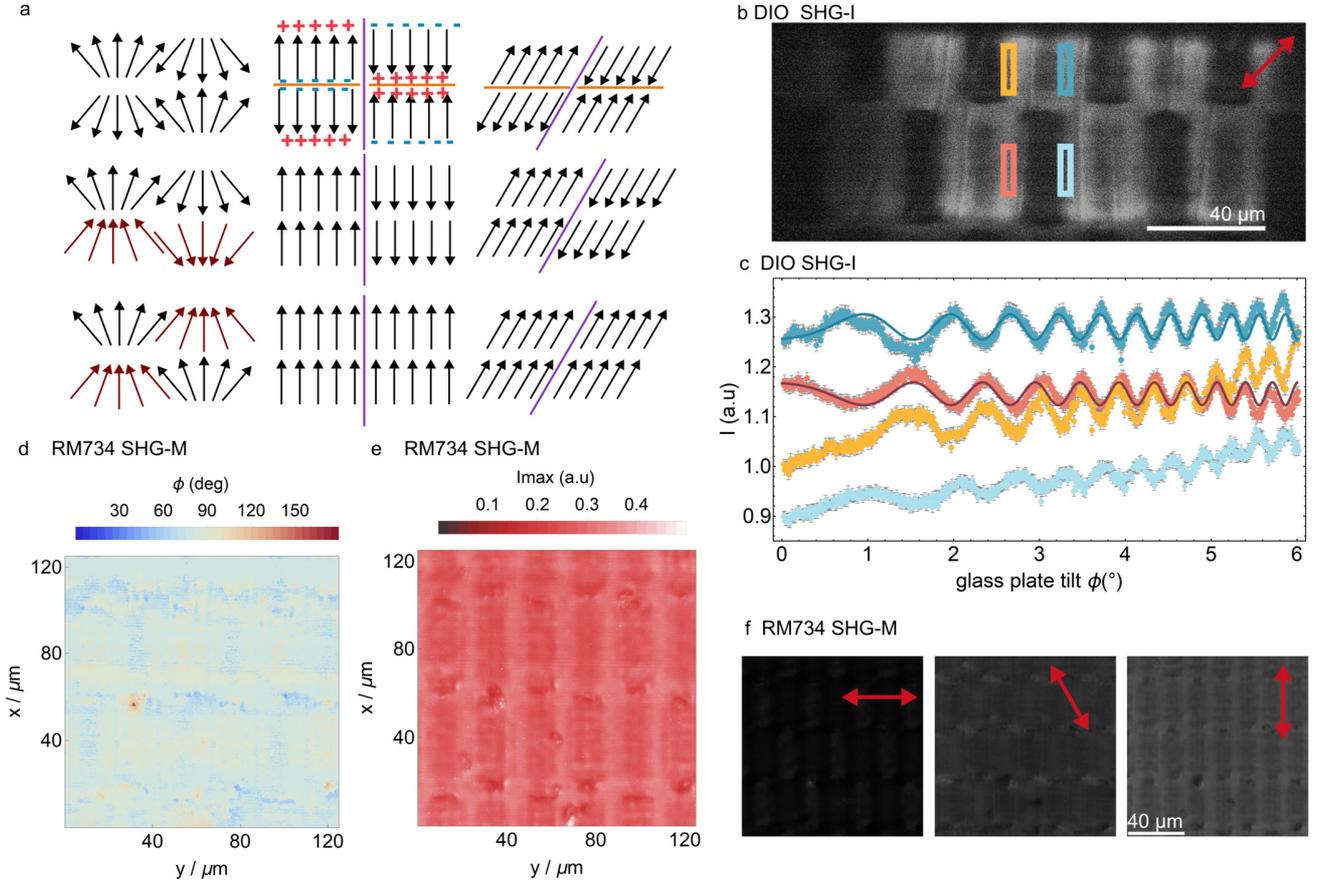

Fig.2. Second Harmonic Generation Microscopy and Interferometry in pattern A of DIO and RM734 for thin cells, i.e. 3 μm cells. a) Schematics of the different possible weakly (purple lines) and strongly charged domain walls (orange lines), showing the orientation in the surfaces and in the middle of the cell for both, non-twisted and twisted areas. Maroon arrows indicate an unfavourable splay/polarization combination. b) SHG interferometry image and c) corresponding interferogram for the considered 2D pattern for DIO. The white arrow indicates the direction of the analyzer. Data in c corresponds to the highlighted areas in b), where solid lines are fits according to Supplementary Eq. 3. d) SHG microscopy images of pattern A for RM734 at different incoming pump polarization (red arrows).

If we focus on the RM734 pattern, the recorded SHG intensity dependence on the incoming pump laser polarization shows that despite the patterned structure, the maximum intensity is obtained across the motif for incoming polarization along the y-direction, i.e. along the average director direction (Fig.2.). The effect of the patterned splay structure is only detected by the slightly larger intensity obtained in those lines in which the director is uniform across the thickness, as opposed to the regions where $\vartheta_{surf}$ deviates from y-direction. The situation differs in the case of DIO. First, the reported $d_{33}$ value for DIO (0.24 pm V$^{-1}$) [11] is an order of magnitude lower than that for RM734 (5.6 pm V$^{-1}$) [10], leading to an overall lower intensity. Second, while for RM734 the $d_{33}$ coefficient is predominant, for DIO the $d_{15}$ coefficient also contributes notably to the signal [26]. Although the pattern exhibits a certain degree of dependence on the incoming pump polarization at which different areas have their maximum SHG signal, the overall lower signal reduces the contrast (Fig.SI.6), making them not so promising for high performance in applications.

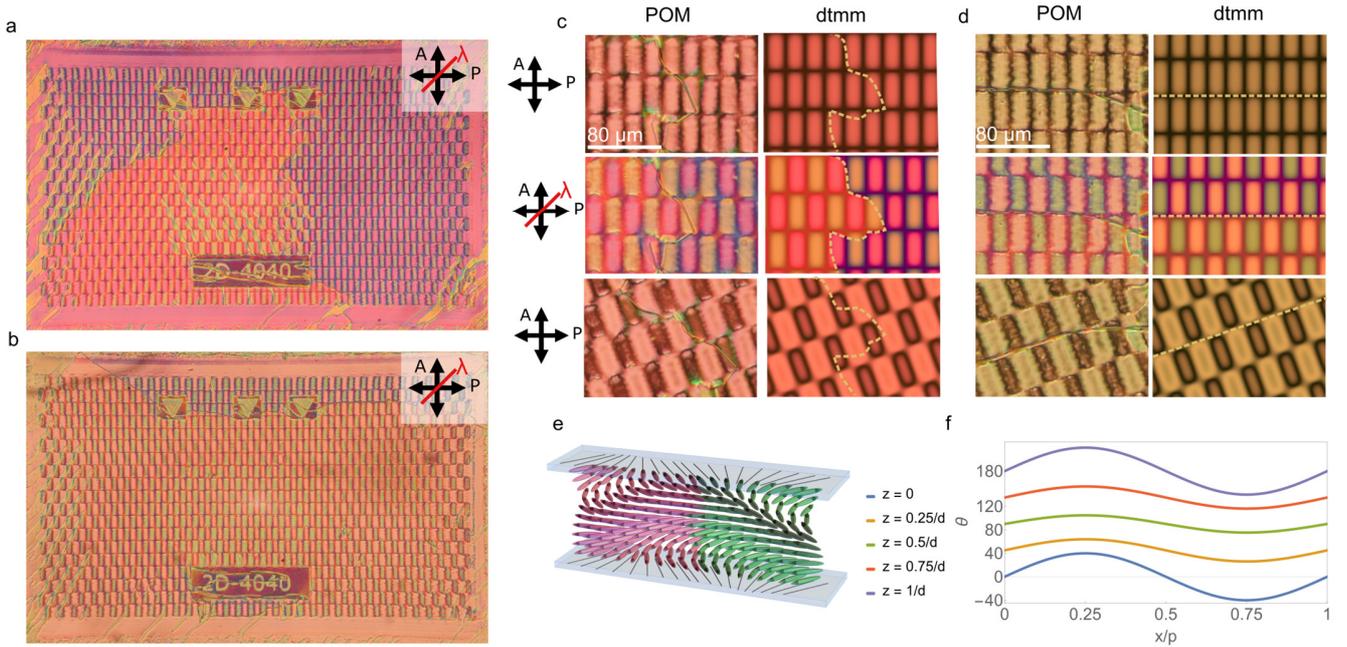

Fig.3. RM734 in pattern A for cells with thickness 8.1 and 10.1 $\mu$m. a&b) Overview of the complete patterned area between crossed polarizers and with the lambda plate for the 8.1 and 10.1 $\mu$m cells, respectively. c) Comparison for three geometries of the POM observations and dtmm simulations considering a $\pi$-twisted structure across the wall dividing two domains of opposite handedness for the 8.1 $\mu$m cell. d) Comparison for three geometries of the POM observations and dtmm simulations considering a $\pi$-twisted structure across the wall dividing two domains of opposite handedness for the 10.1 $\mu$m cell. In both cases $\xi = 1$ is considered. e) 3D sketch of the director structure used in dtmm simulations, where colors denote regions with opposite splay prescribed by anchoring and f) the corresponding $\vartheta$ profiles along x-direction at different levels through the cell thickness for $\xi = 1$.

### 3.2 Pattern A: Splay 2D pattern in thicker cells

We additionally investigated the effect of cell thickness in the final director structure across *d*. For d=5.4 $\mu$m cell DIO adopts the same 2D unsplay structure as can be deduced from the comparison of POM observations and dtmm simulations (Fig.S.7), and thus we will just focus on RM734. Fig.3 depicts the full motif as observed between crossed polarizers and with a full wave plate inserted for RM734 in 8.1 $\mu$m and 10.1 $\mu$m thick cells. In both cases, two regions spanning through a large area of the full design with different optical behaviour can be clearly distinguished. In both cases crossed-polarizers optical observations can be well reproduced considering a $\pi$-twist across $d$ in addition to the surface inscribed pattern, i.e. $\vartheta(z) = \vartheta_{surf} \pm \pi z/d$ (Fig.3.e-f), where the difference between both regions arises from the twist handedness. It should be noted that, as shown by simulations, POM images in these twisted structures hinder obtaining information about the degree of unsplay across the thickness, as structures considering from $\xi = 10$ to $\xi = 0.2$ result in similar transmission patterns compatible with the observations (Fig.SI.8). The simulations confined in Fig.3 were done considering $\xi = 1$. However, the overall twisted structure reduces electrostatic energy, compensating for the depolarization field and the increased elastic energy, as originally proposed by Khachaturyan who considered the electrostatic self-energy of **P**, indicating that a ferroelectric fluid is unstable against twist perpendicular to **P** [9,34].

The SHG-M observations show that through such a combination of surface 2D splay anchoring direction and the helicoidal twist we can produce 2D SHG active patterned arrays, in which adjacent domains exhibit maximum generated SHG signal for incoming pump polarizations at different angles. Results are summarized in Fig.4 for both thicknesses and, in addition to the pattern with $\vartheta_0$ = 40 degrees and P=2$\pi$/k= 40 $\mu$m (A40P40), for a pattern in which $\vartheta_0$ = 60 degrees and P=2$\pi$/k= 60 $\mu$m (A60P60). The analysis of the angle dependence of the intensity throughout the pattern shows that the maximum intensity is obtained at 40 and 120 degrees (where 90 degrees correspond to the vertical direction in the images) for the A40P40 patterns, with only a slight difference depending on the cell thickness, while for A60P60 the angles correspond to 20 and 140 degrees. Such features are well reproduced in SHG simulations (Fig.4) considering the same structure used in Fig.3 for POM. A slight asymmetry can also be observed in the intensity profile through the motif. This observation is interesting as, in order to reproduce it in simulations, a certain degree of unsplay is required close to the surface so as the twist profile slightly varies throughout the pattern. Patterns considering a perfect linear twist throughout the motif (i.e. $\xi = 10$) do not reproduce the intensity asymmetries. It is important to note here that simulations for these twisted structures showed little dependence on the value of $d_{31}$, which was varied in the range from 0 to 0.1 $d_{33}$.

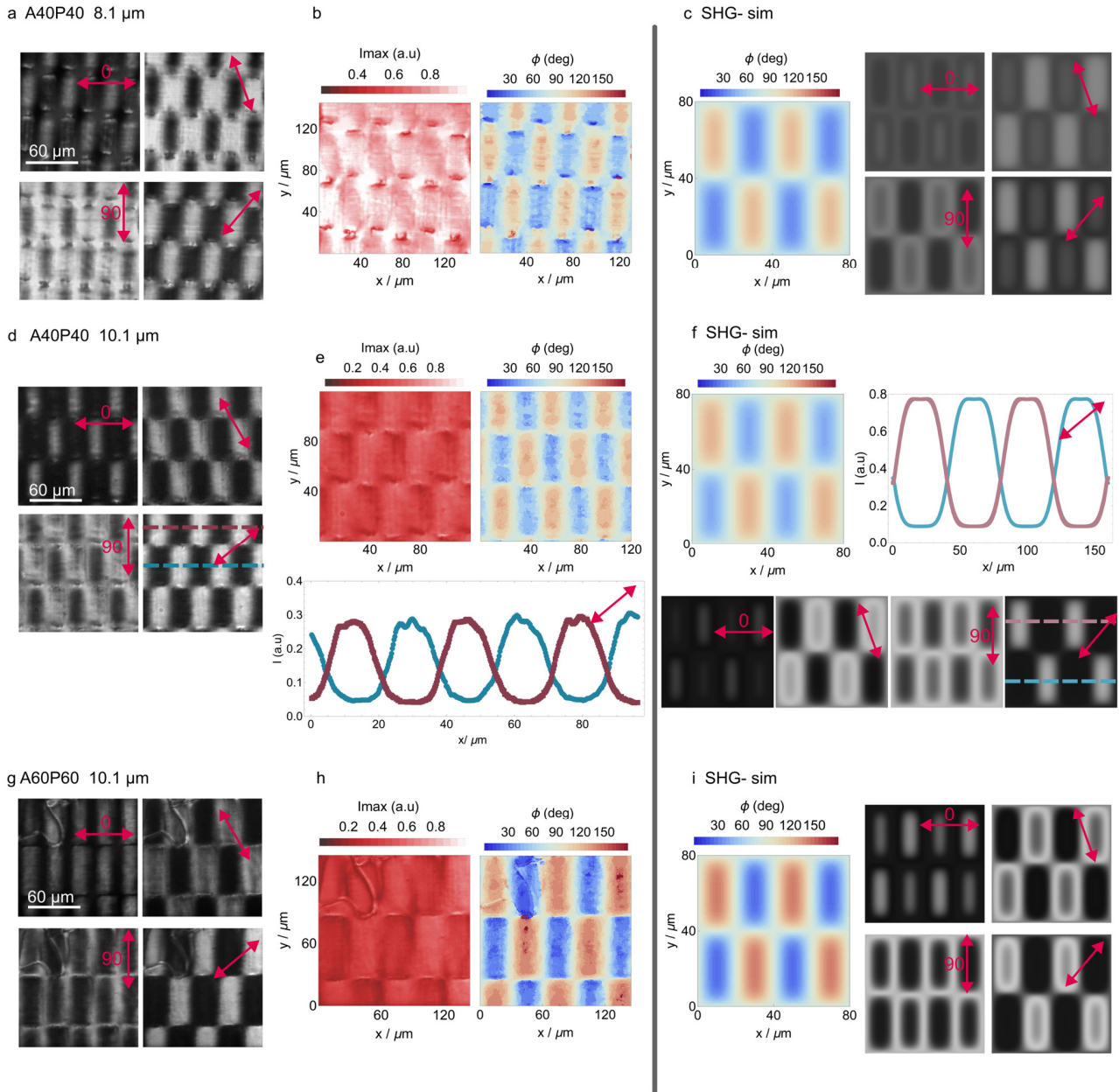

Fig.4. SHG-M and simulations for RM734 in pattern A in 8.1 and 10.1 $\mu m$ thick cells. a) SHG-M images for the A40P40 pattern at different incoming pump polarization directions for 8.1 $\mu m$ cells. b) Analysis of the maximum SHG intensity and corresponding direction of the pump polarization (90 degrees correspond to vertical direction). c) SHG simulated images for the A4040 pattern at different polarizations of the pump beam and corresponding analysis showing the angle at which maximum intensity is obtained. d-f) Equivalent experimental results and simulations as those depicted in a-c, for the same structure in a 10.1 $\mu m$ cell. Panel (e) includes the intensity profile of the 2D pattern for incoming polarization at 120 degrees through the cuts indicated in d by matching dashed lines. Same analysis is included for simulations in panel f. g-i) Equivalent experimental results and simulations as those depicted in a-c, for the A60P60 pattern in a 10.1 $\mu m$ cell. Double headed arrows denote the direction of the incoming polarization.

Inspired by these results, we investigated the effect of sample thickness on the expected SHG signal and angle for the maximum signal for the considered $\pi$ twisted structure and for a structure with $2\pi$ linear twist across the cell. Results for the A40P40 and A60P60 patterns are given in Fig.SI.9. The maximum SHG intensity at each of the complementary areas of the patterns follows a periodic dependence for both $\pi$ and $2\pi$ twisted structures, with the maximum and minimum positions being shifted by half a period between both twists. In the thickness regions in which the signal is near the maximum, little variations are observed in terms of the incoming polarization angles for which the maximum signal is obtained in both adjacent areas. Around the minima however, larger deviations are obtained.

The results confined in Fig.4. together with the performed simulations show that the combination of twist and surface varying angles is a promising avenue for achieving 2D SHG active arrays from polar nematic systems.

### 3.2 Pattern B: Splay 2D including bend

The larger NLO coefficients of RM734 make it a more attractive candidate than DIO to explore SHG patterns. However, the high operating temperature impedes its direct implementation for future applications. Thus, this second pattern design was tested and compared for two materials, RM734 and FNLC-1571, as the latter exhibits $N_F$ at room temperature. In addition, for both materials the main component of the second-order dielectric susceptibility tensor is $d_{33}$ (Fig.SI.12), with comparable values.

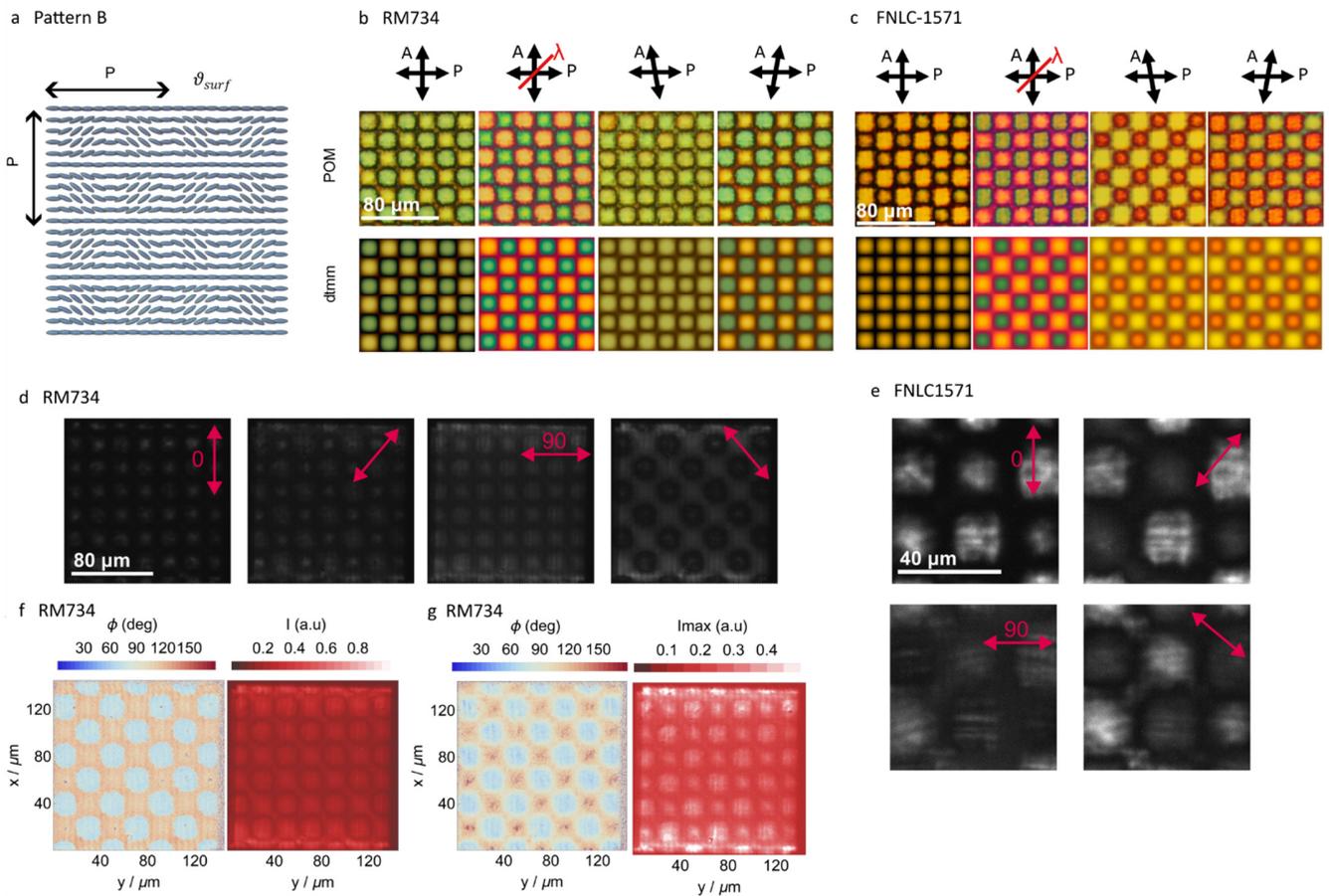

Fig. 5. Pattern B. a) Sketch of the photopatterned design inscribed in both cell´s surfaces. b&c) POM images at different conditions (cross polarizers, cross polarizers in combination with a full-wave plate, analyzer uncrossed 20 degrees clock- and anti-clockwise) are compared with dtmm transmission spectra simulations for RM734 (b) at 120 ℃ and FNLC-1571 (c) at room temperature. d) SHG-M images for RM734 recorded for different pump polarizations, indicated by the double headed arrow. e) SHG-M images for FNLC-1571 recorded for different pump polarizations. f) Analysis of the angle of incoming polarization for which maximum SHG signal is obtained (left) and the corresponding maximum intensity. Analysis is done pixel by pixel. g) Equivalent analysis as for (f) but for experiments performed using an analyzer along the main direction of the pattern, i.e. horizontal.

We first investigated the behaviour of FNLC-1571 in 1D splay patterns comparable to those already explored for DIO and RM734 [26]. Equivalently to what was reported in our previous work, 1D motifs result in domains of alternating polarization divided by domain walls placed in those positions in the pattern where splay changes sign. (Fig.SI.10) Investigation of the effect of splay curvature on the patterning success reveals a behaviour closer to that of DIO, indicating that due to the FNLC-1571 characteristics (i.e. flexoelectric coefficients and spontaneous polarization value), the ideal splay curvature is, in this case, larger than in the case of RM734 (Fig.SI.11).

With this basic characterization, we then explored 2D patterns as those depicted in Fig. 5. In this case, sinusoidal modulation was additionally added in y-direction in order to obtain smooth variation in both directions of the splay sign, i.e $\vartheta_{surf} = A \sin[2\pi x/P] * \sin[2\pi y/P]$, with A=45° and P either 40 or 60 $\mu m$. Such modulation additionally introduces bend deformation in the structure. Periodicities of 40 and 60 μm with a maximum angle of 45 degrees were explored. Due to the layout of the photopatterned cells, for each material, two equivalent patterns were investigated. Interestingly, two kinds of behaviour are observed. While one of the patterns is practically defect-free, the other shows a scramble of domain walls running along y-direction, undulating according to the pattern's bend (Fig.SI.13). Such opposed behaviour speaks for the instability of the structure for this design, in which details of small thickness gradients, inhomogeneities and surrounding unpatterned areas can tilt the scales towards one or another structure. Here we analyze some details of the defect-free pattern.

In the case of FNLC-1571, the patterns throughout the area mostly show symmetric behaviour under crossed polarizers, with asymmetries only observed close to the edges of the pattern (Fig.SI.14). POM observations under different geometries are well reproduced by dtmm simulations considering again a structure that twists across the cell thickness in order to achieve a uniform director orientation (Fig.5 and Fig.SI.14). This "unsplay" is again well reproduced by $\xi = 0.2$. The slight asymmetry observed in the case of RM734 can be reproduced considering a small deviation in the middle of the cell $\vartheta_{center} = 5°$ while in those asymmetric areas in FNLC-1571 $\vartheta_{center}$ in found to be closer to 15°. With this in mind, for SHG experiments in FNLC-1571 we will focus on the center of the motif where symmetric behaviour is observed.

SHG microscopy images for both materials in pattern B are shown in Fig.5 for different incoming pump polarization directions. Fig.5.f&g. shows the analysis of the maximum SHG recorded signal throughout the pattern together with the incoming pump polarization direction at which such maximum intensity is recorded for RM734. Although a clearly distinguishable angle dependence can be observed, the intensity contrast between adjacent areas is small. Notably, the analysed area for RM734 shows a sharp change of incoming pump polarization angle for maximum intensity in measurements performed without analyser, while a smooth variation is observed employing an analyser. Such observation speaks for the importance of NLO coefficients other than $d_{33}$ for the design of SHG active arrays. It should be noted here that it was found that SHG simulations for such untwisted structures are drastically sensitive to the combination of input dispersion, cell thickness and NLO coefficient values in comparison to the twisted structures, and thus can show misleading results in the range of $d_{31}=d_{15}=0$ and $d_{31}=d_{15}=0.1d_{33}$.

It is interesting to note here, that FNLC-1571 was maintained at room temperature ($N_F$) for a year, after which optical and SHG experiments were repeated for assessing the stability of the structure. Over the course of a year, the monodomain pattern remained stable, without decaying in multiple domains. However, the structure throughout the whole motif relaxed into the structure with asymmetric behaviour, with the director pointing uniformly at around 12 degrees in the centre of the cell (Fig.SI.15). Consequently, SHG signal profile across the pattern was maintained, with just a slight shift in the angular behaviour with respect to the original state (Fig.SI.15c).

Summarizing for the pattern B in which the motif is based on smooth modulation in both directions, which in addition to splay deformation involves bend, a clear dependence on the pump polarization angle with respect to the pattern is obtained for adjacent areas. However, the inclusion of bend seems to strongly destabilize the structure, in the way that equal patterns can results either in large structured monodomains or in the decay in multiple domains.

## 4. Conclusions

To conclude, we have demonstrated the pathway towards the fabrication of tailored NLO arrays in the micrometre range, by exploiting patterned photoalignment possibilities. We show that a balance between flexoelectric and electrostatic energy minimization can be used to create a number of structures. However, one should note that material parameters (i.e. spontaneous polarization, ideal splay curvature) combined with design parameters (i.e. cell thickness, periodicity and amplitude of the splay patters) all impact the final achieved structure. In addition, the NLO coefficients together with the dispersion of refractive indices for the target wavelengths also determine the final SHG operational behaviour of the motifs.

For the two investigated patterns based on simple splay structures, we show that in order to minimize electrostatic energy and avoid the creation of strongly charged domain walls, the different materials adapt to unfavourable splay in different ways. Notably, for RM734 large monodomain patterns are obtained with the director at the surfaces following the prescribed alignment and rapidly twisting towards a uniform direction in the middle of the cell. This represents an advantageous scenario when increasing the cell thickness, as such surface structure is shown to procure the decay of the pattern into two 180 degrees twisted domains of opposite handedness. Such twisted structures are shown to be highly profitable from the perspective of achieving 2D SHG active arrays showing strong contrast between adjacent areas for opposite angles. Considering the importance of the twisted structures for the final SHG patterning, a very promising avenue is to combine such surface boundary conditions to ferroelectric cholesteric nematics, for which phase-matched SHG near the photonic bandgap has been recently reported [20,21]. It is also interesting to note here, that precise control and alignment of $N^*_F$ is not only promising for creating photonic structures, but also for the design of $N^*_F$ based lasers with enhanced lasing properties and optimized tunability [35,36].

It should be noted that the design optimization will greatly benefit from a detailed model for ferroelectric nematic materials, allowing to input the material parameters to predict energy minimizing structures. In combination with accurate knowledge of materials parameters and SHG simulations, such a model would greatly expedite the implementation of ferroelectric nematics as key photonic materials.

## Declaration of Competing Interest

The authors declare no conflict of interest.

## Acknowledgements

The ferroelectric nematic liquid crystal FNLC-1571 used in this work was supplied by Merck Electronics KGaA. N.S, A.M, I.D-O., M.L, N.O, and A.P acknowledge the support of the Slovenian Research Agency (grant numbers P1-0192, N1-0195, J1-50004 and PR-11214-1). K.N and B.B would like to acknowledge the support of the Research Foundation—Flanders (FWO) through grant number G0C2121N. R.J.M. acknowledges funding from UKRI via a Future Leaders Fellowship, grant No. MR/W006391/1. S.A. and M.H. acknowledge the National Key Research and Development Program of China (No. 2022YFA1405000) and the Recruitment Program of Guangdong (No. 2016ZT06C322).

## Data availability

Data will be made available upon reasonable request. Code for diffractive transfer matrix method (dtmm) optical simulations is deposited in Github/Zenodo at "Andrej Petelin. (2020). IJSComplexMatter/dtmm: Version 0.6.1 (V0.6.1)", under accession code https://doi.org/10.5281/zenodo.4266242.

# Supplementary Material

# Patterning of 2D second harmonic generation active arrays in ferroelectric nematic fluids.

M. Lovšin[1,2], A. Petelin[2], B. Berteloot[3], N. Osterman[1,2], S. Aya[4,5], M. Huang[4,5], I. Drevenšek-Olenik[1,2], R. J. Mandle[6,7], K. Neyts[3,8], A. Mertelj[1], N. Sebastian[1]

1 Jožef Stefan Institute, Ljubljana, Slovenia
2 University of Ljubljana, Faculty of Mathematics and Physics, Ljubljana, Slovenia
3 Liquid Crystals and Photonics Group, ELIS Department, Ghent University, Ghent, Belgium
4 South China Advanced Institute for Soft Matter Science and Technology (AISMST), School of Emergent Soft Matter, South China University of Technology, Guangzhou, China
5 Guangdong Provincial Key Laboratory of Functional and Intelligent Hybrid Materials and Devices, South China University of Technology, Guangzhou, China
6 School of Physics and Astronomy, University of Leeds, Leeds, UK
7 School of Chemistry, University of Leeds, Leeds, UK
8 State Key Lab of Advanced Displays and Optoelectronics Technologies, Hong Kong University of Science and Technology, Hong Kong
Corresponding author: Nerea.sebastian@ijs.si

## Index

# Supplementary Note I – Determination of dispersion for RM734

Dispersion value for the extraordinary refractive index used in SHG simulations was determined for RM734 using the wedge-cell method. For that a EHC KCRS-03 wedge cell with $\tan(\varphi) = 0.0115$ and parallel rubbing along the wedge was filled with RM734. Using the setup described in [1] SHG microscopy images were taken at different positions clearly showing Maker's fringes and as a function of temperature. To determine dispersion data, it is considered that the SHG intensity at normal incidence depends on the thickness L as $\sin^2(\Delta k\, L/2)$, where

$$\Delta k = k_{2\omega} - 2k_\omega = \frac{2\pi}{\lambda/2}n_{2\omega} - \frac{4\pi}{\lambda}n_\omega$$

Being $\lambda$ the wavelength of the incident light ($\lambda$=800 nm in our case). Considering two consecutive SHG maxima we have that $\Delta k = 2\pi/\Delta L$ and then dispersion can be obtained as $n_{2\omega} - n_\omega = \lambda/2\Delta L$. Calculating $\Delta L$ as the average distance between intensity peaks at a given temperature (Fig.SI.1), obtained dispersion data for the extraordinary indices of refraction of RM734 is given in Fig.SI.1 as a function of temperature.

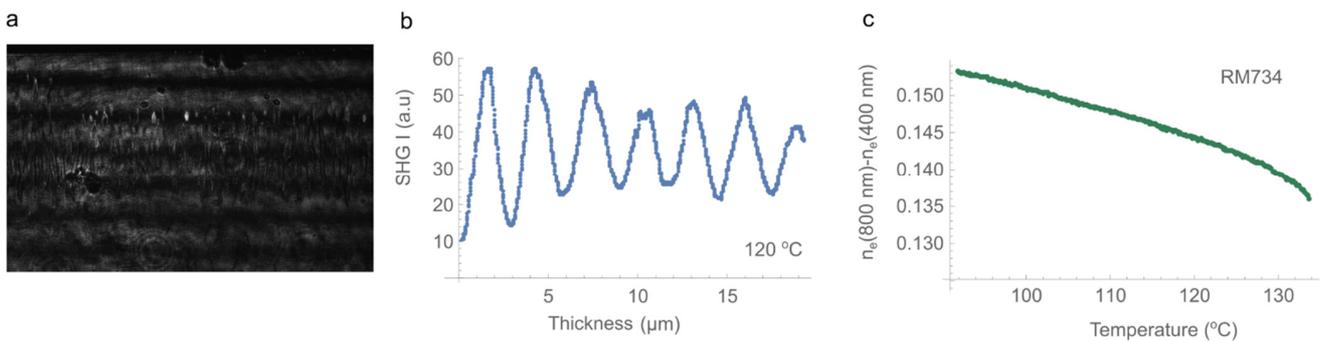

Fig.SI.1 Determination of dispersion data for the extraordinary indices of refraction of RM734. a) SHG microscopy image in the thinner part of the wedge cell corresponding to thicknesses between 0 and 20 μm. b) Corresponding SHG intensity profile. c) dispersion data for the extraordinary indices of refraction $n_{2\omega} - n_\omega$ of RM734 for the wavelengths $\lambda = 800$ nm and $\lambda = 400$ nm

# Supplementary Note II – Pattern A

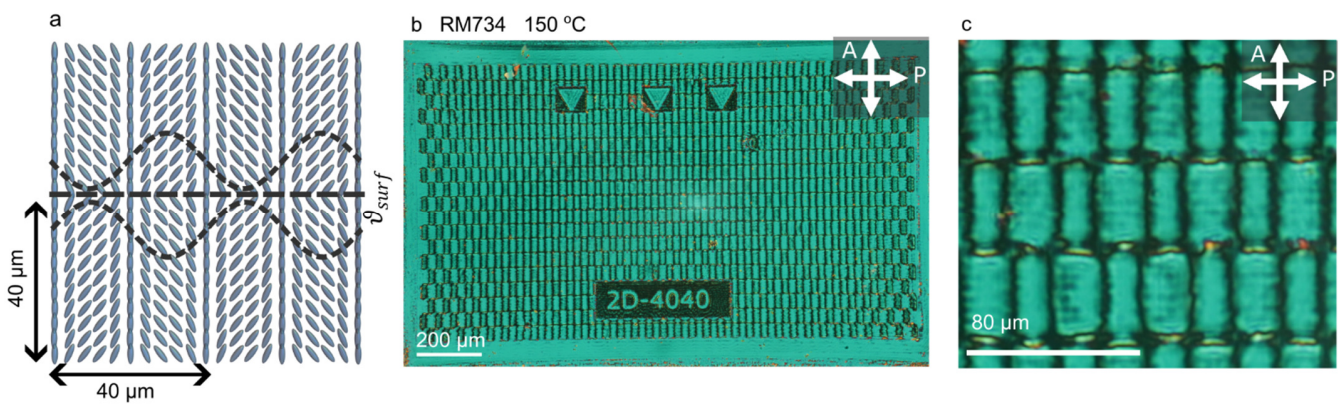

Fig.SI.2 2D pure splay patterns in RM734 in the high temperature non-polar nematic phase. a) Photopatterned design inscribed in both cell´s surfaces. b) Full patterned area and c) zoom-in POM images for RM734 at 150 °C under crossed polarizers.

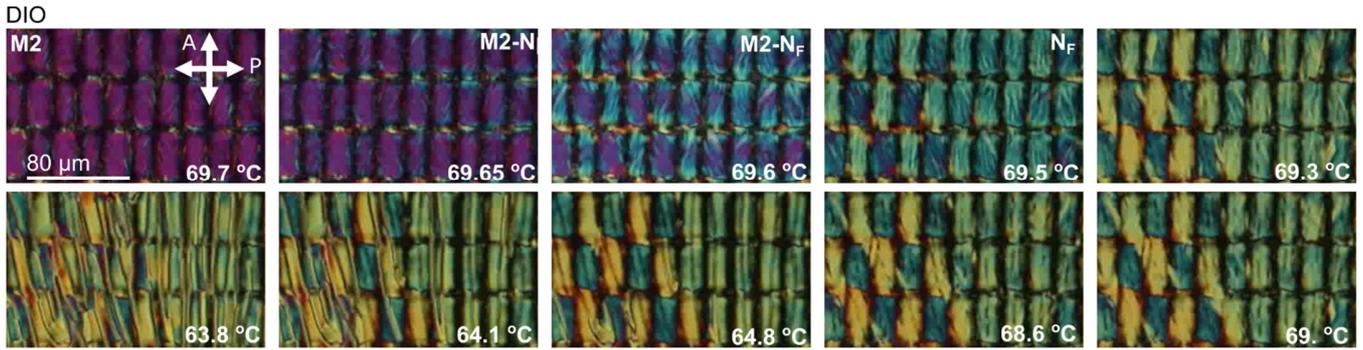

Fig.SI.3 M2-N$_F$ phase transition as observed in the pure splay 2D pattern ($\vartheta_0$ = 40 degrees and the period P=40 μm) for DIO in a 3 μm cell.

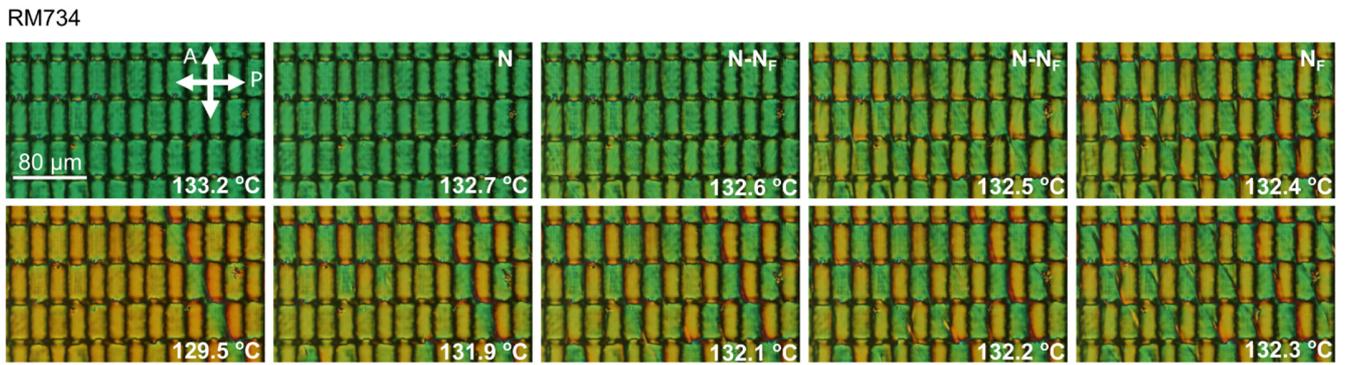

Fig.SI.4 N-N$_F$ phase transition as observed in the pure splay 2D pattern ($\vartheta_0$= 40 degrees and the period P=40 μm) for RM734 in a 3 μm cell.

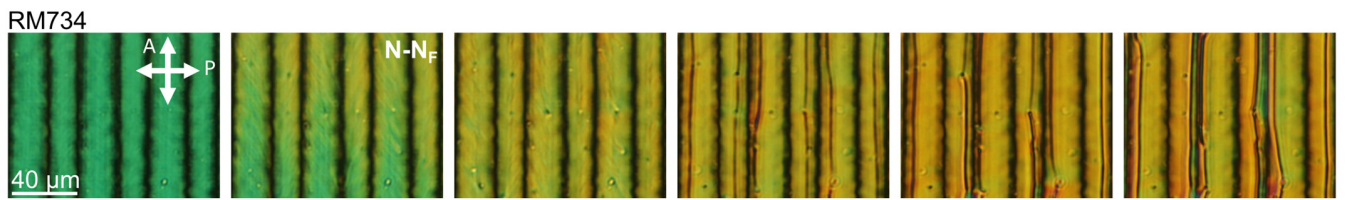

Fig.SI.5 Periodic 1D splay pattern ($\vartheta_0$ = 40 degrees and the period P=40 μm) for RM734 in a 3 μm cell during the N-N$_F$ phase transition, showing the appearance of defect lines across the pattern.

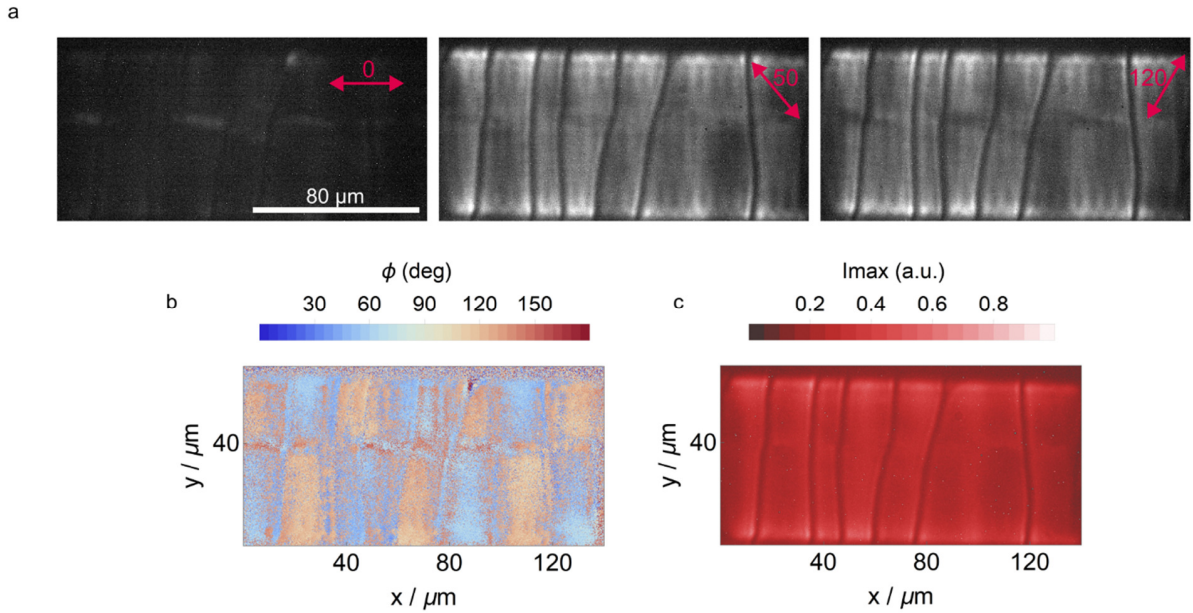

Fig.SI.6 SHG microscopy images of DIO in pattern A for a 3 μm cell. a) SHG-M images for three different incoming pump polarization directions marked with double-headed arrows. b) Polarization direction of the incoming pump for which the maximum SHG signal is obtained and c) the corresponding maximum intensity map throughout the motif. In b, 0 degrees corresponds to horizontal direction.

## Supplementary Note III – Interferograms fitting function

In the SHG interferometry setup, the recorded intensity can be expressed as:

$$I = \left(A_{sample} \sin(kz) + A_{reference} \sin(kz + \delta)\right)^2 \quad (1)$$

The phase between the SHG signal of the sample and that of the reference is varied by inserting a glass slide (thickness d) and varying the incidence angle ($\phi$) and then $\delta$ can be written as:

$$\delta = \delta_0 + \frac{kd}{\sqrt{1 - \frac{\sin^2(\phi)}{n^2}}} \quad (2)$$

where $k = 2\pi n/\lambda$. Interferometry sets of data have then been fitted to

$$I = A_{sample}^2 \left(1 + A \sin\left(\delta_0 + \frac{\frac{2\pi n d}{\lambda}}{\sqrt{1 - \frac{\sin^2(\phi - \phi_0)}{n^2}}}\right)\right)^2 \quad (3)$$

with $\lambda = 780/2$ nm and $A = A_{reference}/A_{sample}$. Reasonable values of n and d for the inserted glass plate were obtained from a set of fits to be n=1.516 and d=1.046 mm and subsequently fixed for the rest of data sets.

# Supplementary Note IV – Pattern A Additional Polarizing Optical Microscopy in thick cells

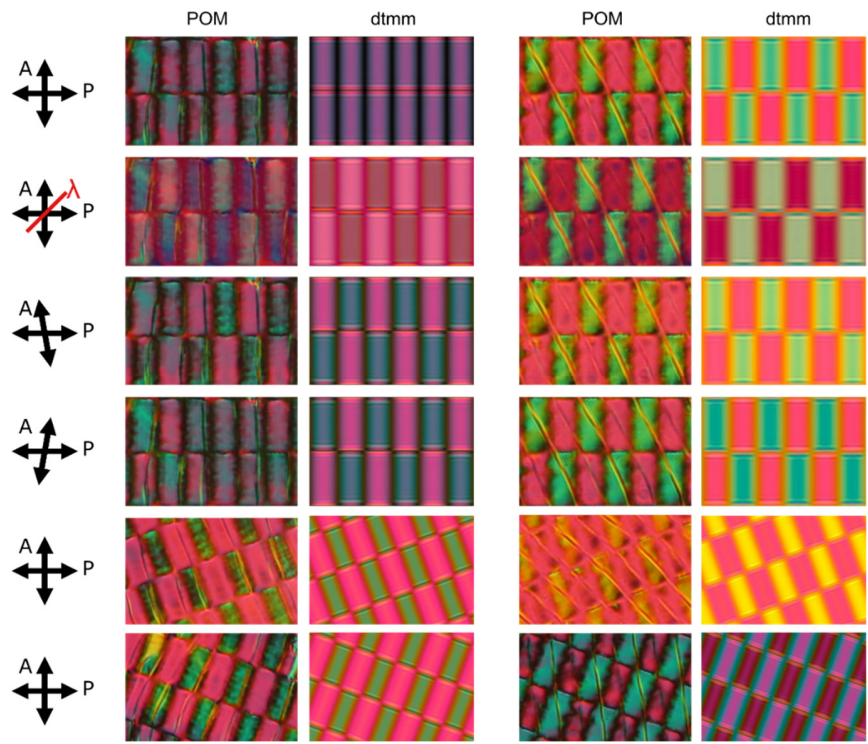

Fig.SI.7 Comparison of POM images and dtmm simulations for DIO and all the studied geometries for the 2D splay pattern with $\vartheta_0 = 40°$, $\xi = 0.2$ and P= 40 $\mu m$ in a 5.4 µm cell. Dtmm simulations were performed considering $\Delta n = 0.19$, where n$_o$ is taken to be 1.52. From top to bottom: Crossed polarizers along the pattern, with lambda plate, uncrossing analyser in opposite directions and rotating the sample in opposite directions. Left and right columns correspond to two different areas of the pattern where symmetric and asymmetric behaviour is observed respectively.

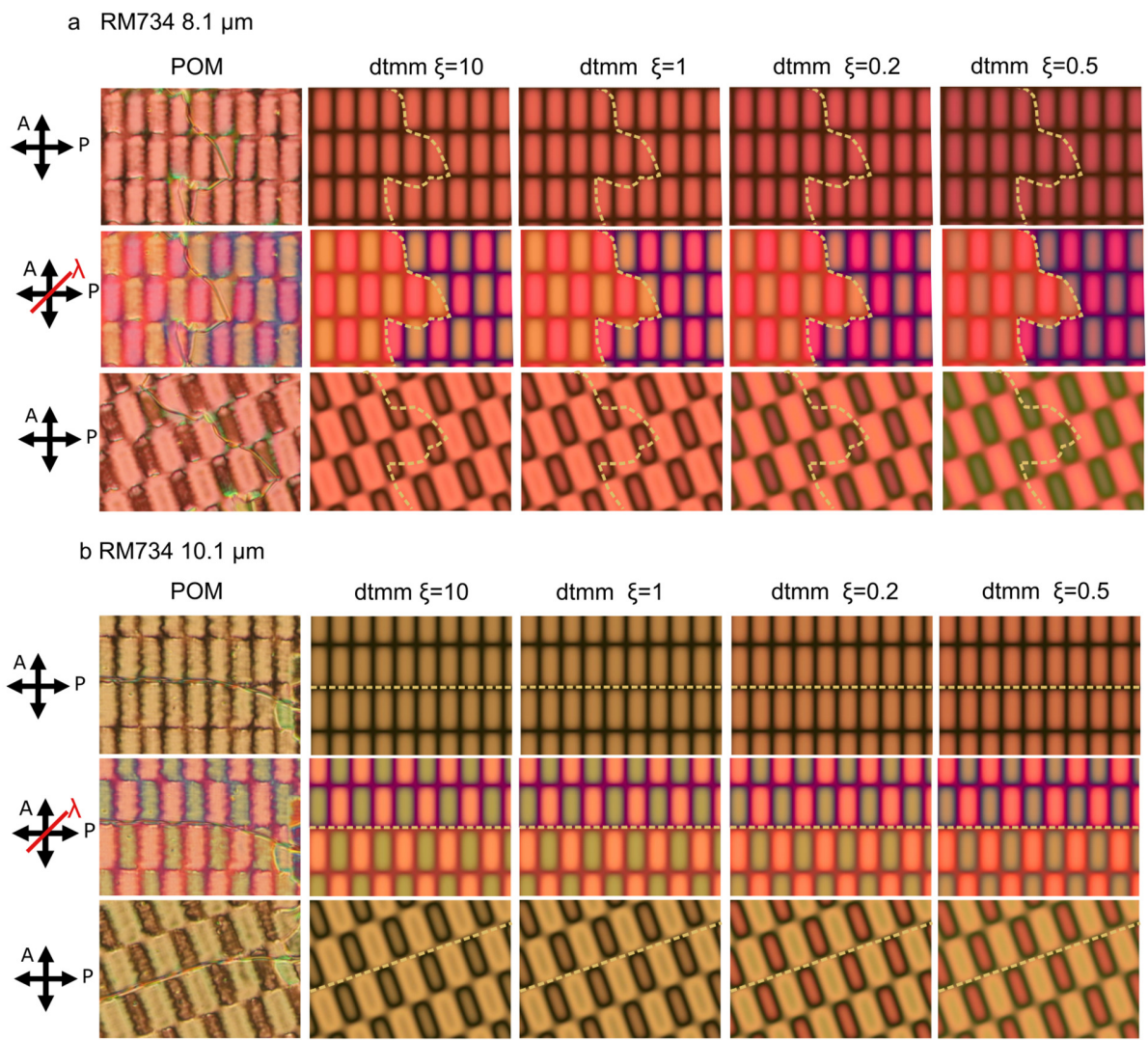

Fig.SI.8 Comparison of POM and dtmm simulations for RM734 in pattern A. a) Comparison for different "unsplay" parameter $\xi$, for RM734 with $\Delta n$=0.25, and d=8.1 μm for extinction position, crossed polarizers and lambda plate and sample rotated for 20 degrees. a) Comparison for different "unsplay" parameter $\xi$, for RM734 with $\Delta n$=0.25 and d=10.1 μm, for extinction position, crossed polarizers and lambda plate and sample rotated for 20 degrees.

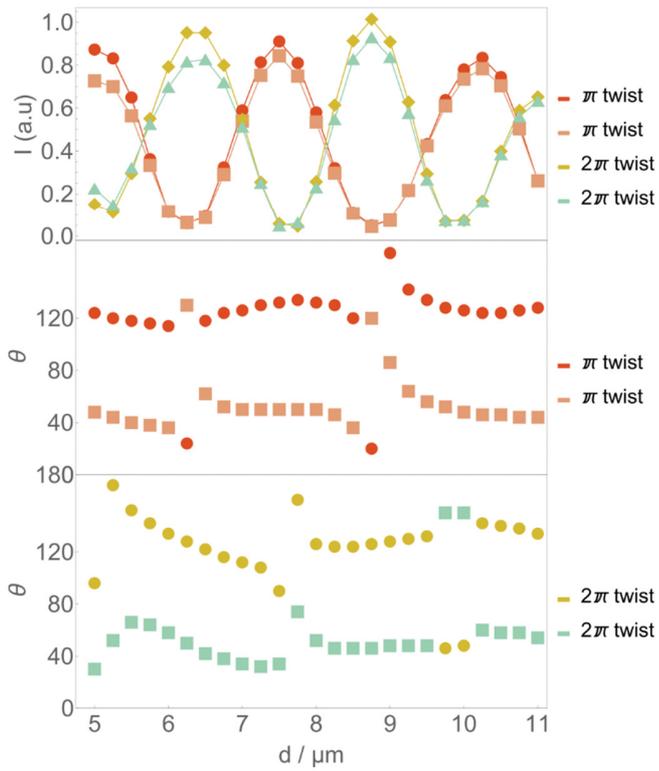
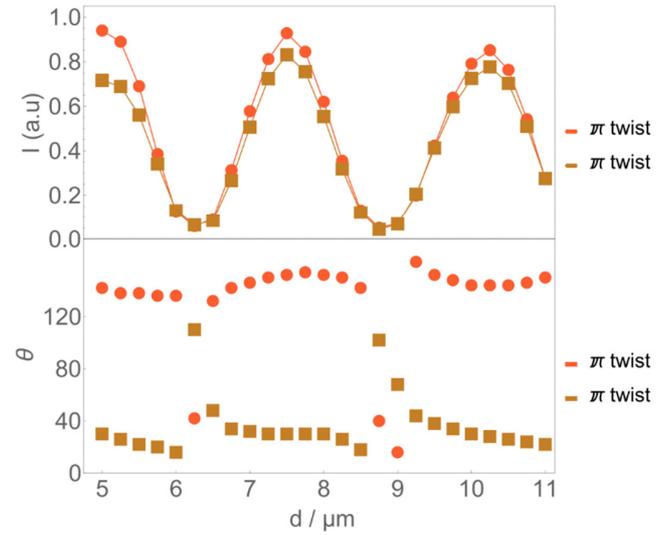

Fig.SI.9 Dependence of maximum SHG intensity and pump polarization angle for maximum intensity on the sample thickness as calculated for A40P40 and A60P60 patterns considering both, linear $\pi$-twist and linear $2\pi$-twist director variation through the cell thickness.

# Supplementary Note V – FNLC-1571 behaviour in 1D splay lines

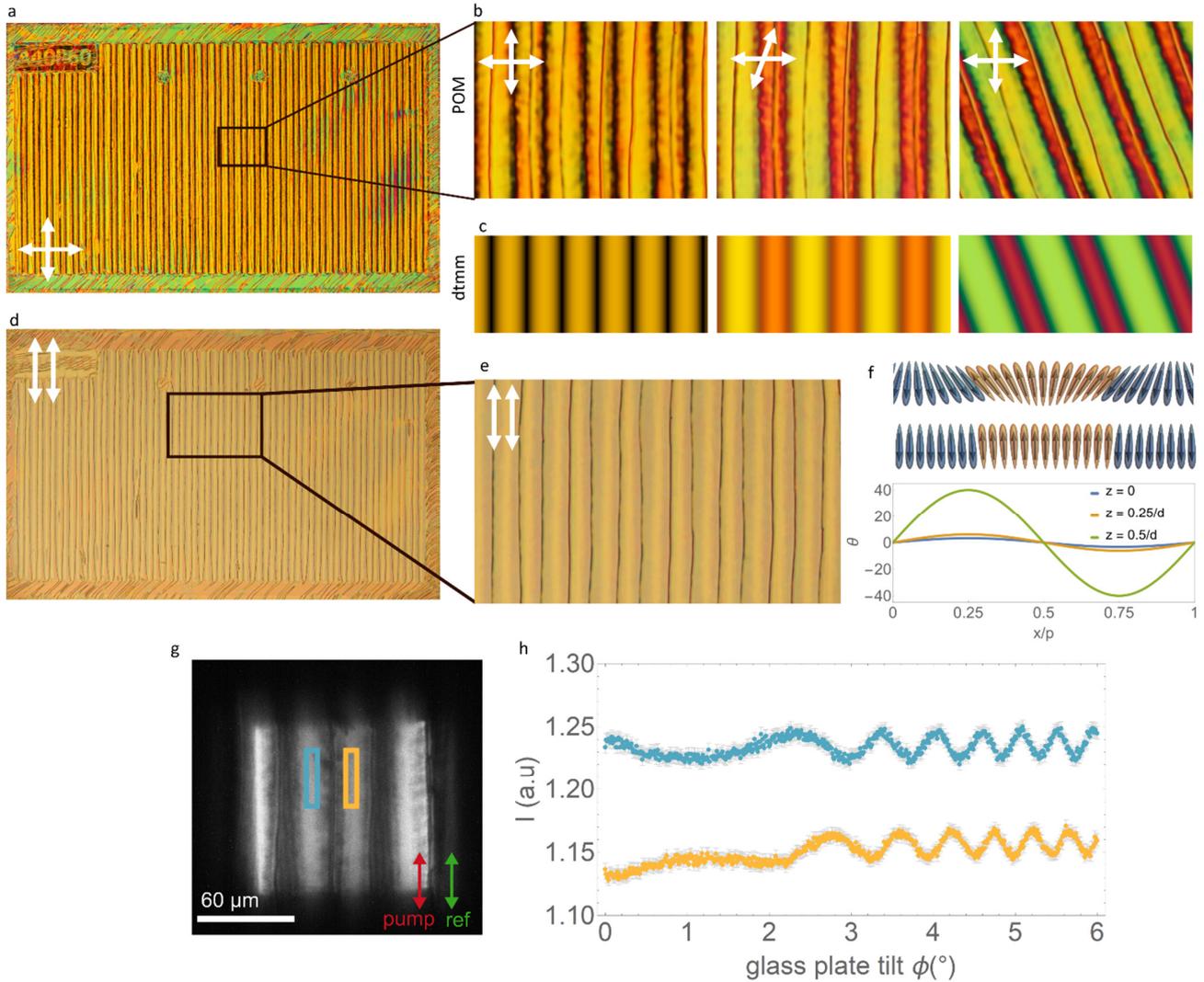

Fig.SI.10 splay modulation pattern for FNLC-1571 in 3 μm cell. a) Overview of the full pattern with $\vartheta_0 = 40$ and periodicity of 60 μm in extinction position. b) Zoom in on the highlighted area in (a) for extinction position, analyzer uncrossed 20 degrees and sample rotated anticlockwise for 20 degrees and c) corresponding dtmm simulations considering the structure $\vartheta(z) = \vartheta_{surf} e^{\left(\frac{2z^2}{d^2} - \frac{1}{2}\right)/\xi}$ with $\xi=0.2$ and $\vartheta_{surf} = 40 Sin(2\pi x/P)$. Values of refractive indexes and dispersion were taken from [2]. g&h) SHG interferometry measurements in two consecutive areas separated by a domain wall. Data in the interferogram corresponds to the SHG intensity variations for the areas highlighted in the image.

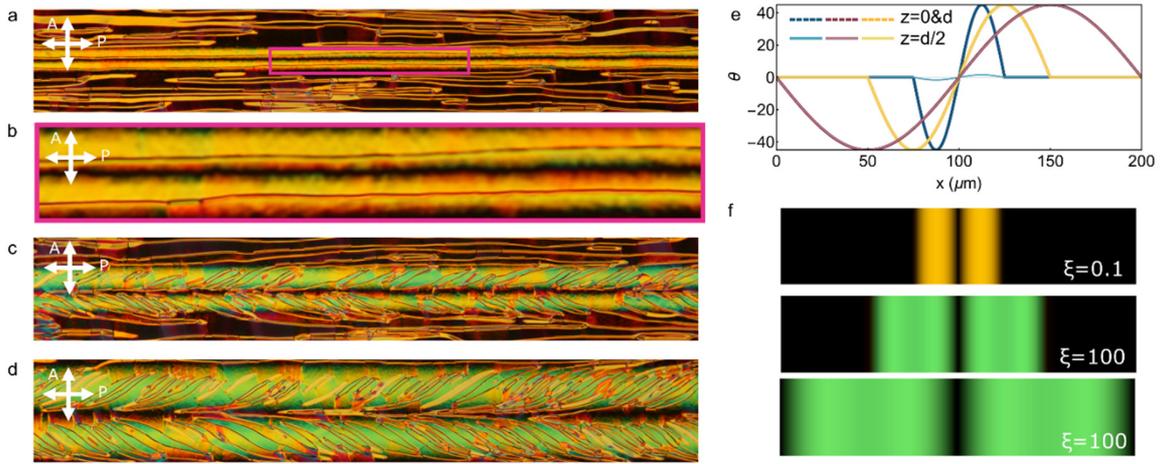

Fig.SI.11 Single splay lines embedded in the uniform background for FNLC-1571 in a 3 µm cell. POM observations of single splay lines with $\vartheta_0=45°$ and (a&b) $P$ = 50, (c) 100, and (d) 200 µm ($k\vartheta_0$=0.1, 0.05, and 0.025 µm$^{-1}$) in a uniform background as observed under crossed polarizers. b) Zoom image of the highlighted area in (a). e) Patterned angle profile (solid lines) and angle profile in the centre of the cell (dashed lines) as deduced form the comparison of POM experiments and dtmm simulations. f) dtmm simulations considering the structure $\vartheta(z) = \vartheta_{surf} e^{\left(\frac{2z^2}{d^2}-\frac{1}{2}\right)/\xi}$ with $\vartheta_{surf}$ given in (e) by solid lines.

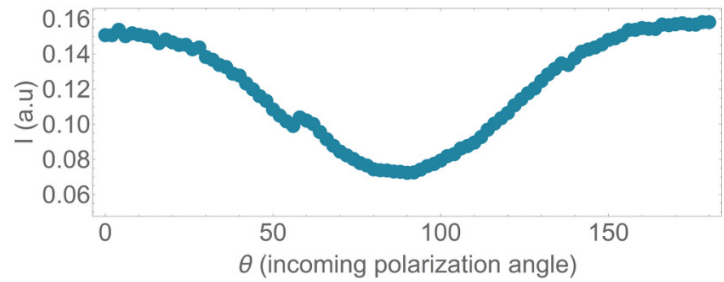

Fig.SI.12 Dependence of the recorded SHG intensity in uniform pattern on the angle θ, measured between the incoming laser polarization and the preferred photoalignment orientation for FNLC-1571.

# Supplementary Note VI – Additional information for Patterns B.

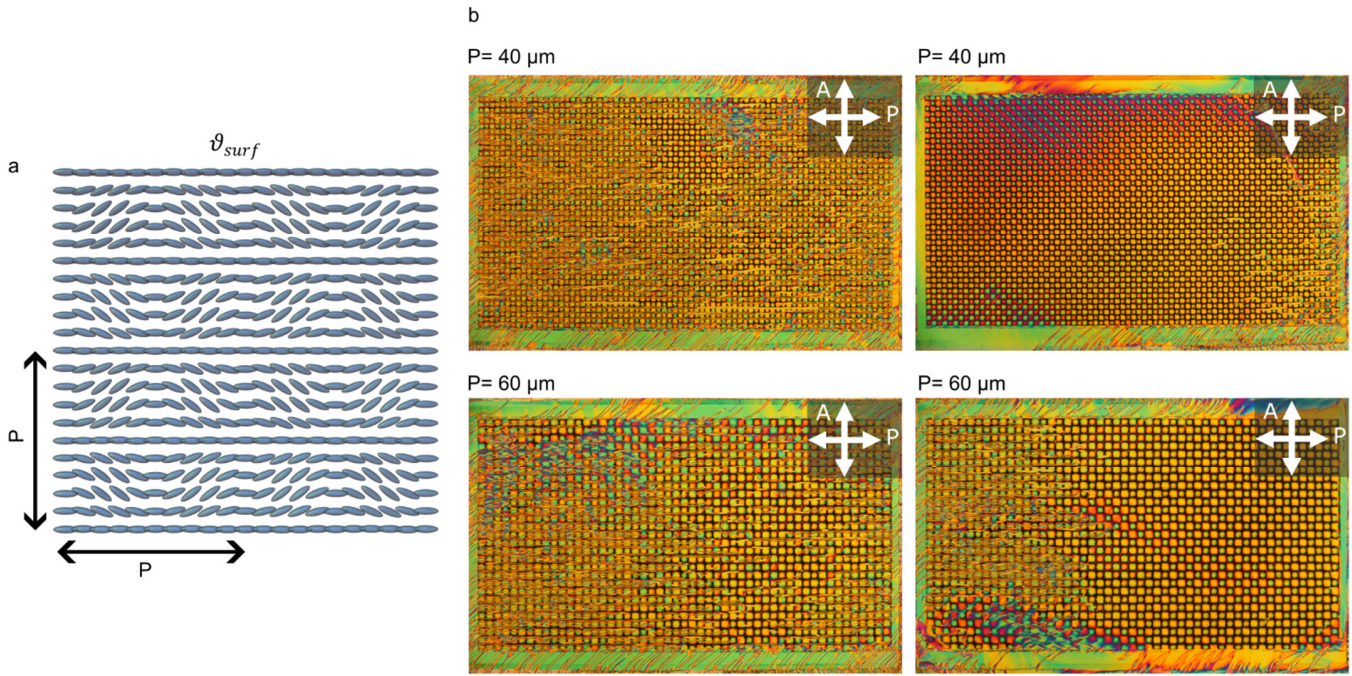

Fig.SI.13 Pattern B. a) Sketch of photopatterned design A $\sin[2\pi x/P] * \sin[2\pi y/P]$, with A=45° and P either 40 or 60 $\mu m$. b) POM image under crossed polarizers (white double-headed arrows) of the full patterned motifs. Two examples for A40P40 and two for A60P60 are given.

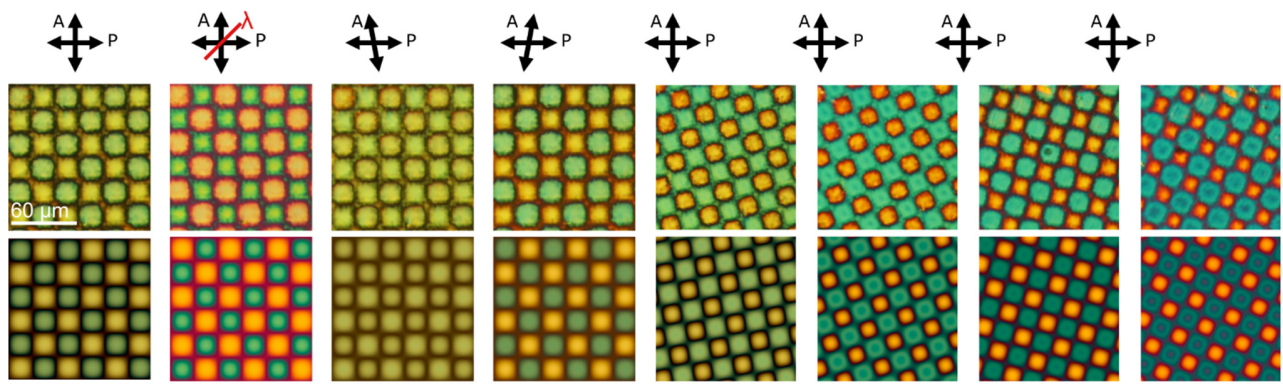
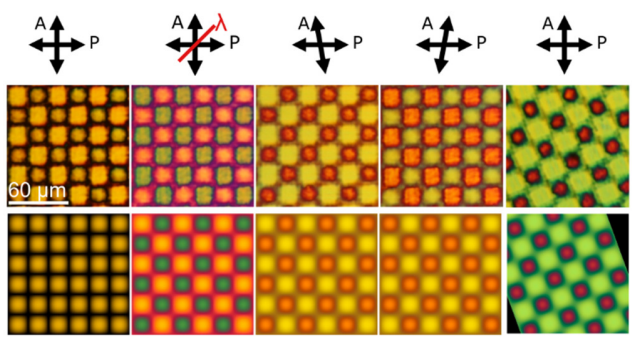
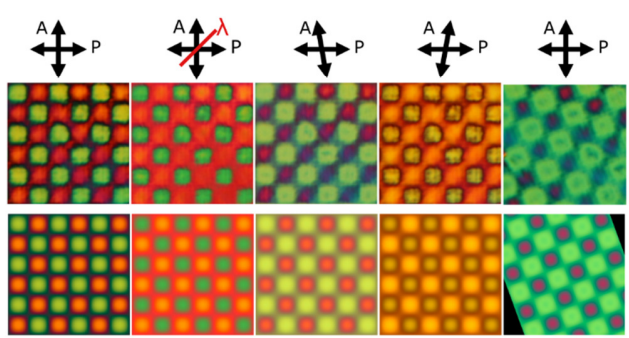

Fig.SI.14 Comparison of POM observations and dtmm simulations at different geometries for pattern B A40P40 for a) RM734, b) FNLC-1571 in the centre of the patterned area (Fig.SI.13.b top-right) and c) FNLC-1571 in the boundary of the photopatterned area (Fig.S.13.b top-right).

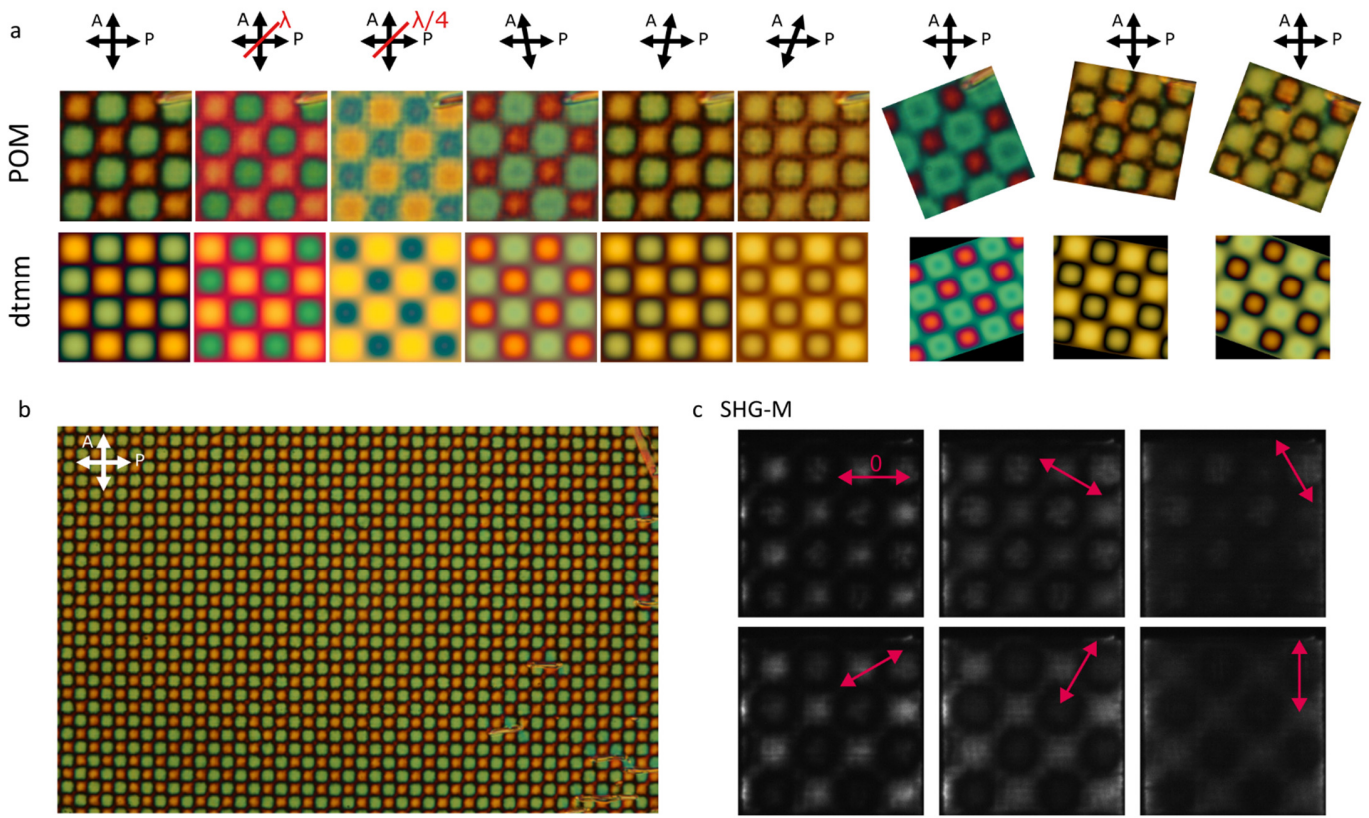

Fig.SI.15 a) Comparison of POM observations and dtmm simulations at different geometries for pattern B A40P40 FNLC-1571 one year after the cell was filled and maintained at room temperature. Simulations were performed considering $\vartheta(z) = \vartheta_{surf} e^{\left(\frac{2z^2}{d^2}-\frac{1}{2}\right)/\xi}$ $+12 e^{\left(\frac{2z^2}{d^2}-\frac{1}{2}\right)/\xi}$ with $\xi=0.2$ and $\vartheta_{surf} = 45 Sin(2\pi x/P) Sin(2\pi y/P)$. b) Overview of the patterned area showing the relaxation of the structure throughout the motif to a structure in which the director twists to become uniform towards the cell centre at 12 degrees with respect to the pattern axis. c) SHG microscopy images at different incoming pump polarizations.